\documentclass[a4paper,11pt]{article}
\pdfoutput=1 

\usepackage{jcappub} 

\usepackage[T1]{fontenc} 
\usepackage{xcolor}
\usepackage{multirow}
\usepackage{soul}

\title{A trium test on beyond $\Lambda$CDM triggering parameters.}

\author[\,a,b,c]{Z. Sakr}

\affiliation[a]{Institut f\"ur Theoretische Physik, Philosophenweg 16, 69120 Heidelberg, Germany}
\affiliation[b]{IRAP, Université de Toulouse, CNRS, CNES, UPS, Toulouse, France}
\affiliation[c]{Faculty of Sciences, Universit\'e St Joseph; Beirut, Lebanon}

\emailAdd{ziad.sakr@net.usj.edu.lb}

\abstract{We performed a Bayesian study on three beyond $\Lambda$CDM phenomenological triggering parameters, the growth index $\gamma$, the dark energy equation of state parameter $\omega$ and the lensing deviation from the GR prediction parameter $\Sigma$, using the latest cosmological geometric, growth and lensing probes, all in a consistent implementation within the modified gravity cosmological solver code MGCLASS. We find, when we combined all our probes, i.e. the cosmic microwave background (CMB), the baryonic acoustic oscilation (BAO), the growth measurements $f\sigma_8$ and the 3$\times$2pt  joint analysis of weak lensing and galaxy clustering in photometric redshift surveys, assuming flat space, constraints compatible with general relativity and $\Lambda$CDM with $\omega = -1.025\pm0.045$, and $\Sigma = 0.992\pm0.022$ at the 68\% level, while $\gamma = 0.633\pm0.044$ is still within $\sim 2 \sigma$ from the $\Lambda$CDM value of $\gamma \sim 0.55$, and that when $\Sigma$ is considered as constant; while $\gamma_\ell = -0.025 \pm0.045$ when the lensing parameter is parameterised as function of a lensing index, introduced for the first time in this work, as $\Sigma(z)=\Omega_m(z)^{\gamma_\ell}$. }

\begin{document}
\maketitle
\flushbottom

\section{Introduction}\label{sec:intro}

The cosmological standard model, dubbed $\Lambda$CDM, is the minimal model in concordance with current observations. The latter however still allows, though within small margins, some extensions to it, such as a different evolution for the dark energy component, or a modification of the general relativity to, either mimic the effect of dark energy (DE) or to test whether new theories could better fit the data along with the presence of a dynamical DE.
We are not going to review here the many attempts to explain the origin of a constant or dynamical dark energy (see \cite{Martin:2012bt} and references therein) or the different modified gravity theories that could or not account for the evolution of the universe in general or its late acceleration in particular (see \cite{Clifton:2011jh} for a review), but we just note that the landscape of possible still viable theories, albeit the fact that they have not been tested enough, has become so extended that it is becoming more difficult to test them exhaustively in a short time with every data release. This is true especially since each model would need substantial work to, first pass the theoretical conditions and not violate some fundamental physical principles, then determine the outcome of the different probes and observables within the new framework, to finally confront them to data throughout time-consuming Monte Carlo Markov Chain (MCMC) Bayesian tests, the most common method commonly used to infer constraints on a specific model's parameters. Therefore many suggested to group or try to encapsulate the different extensions and modifications under parameterized models tailored to test their phenomenological imprints on the cosmological observables. A first simple way was the introduction of the parametrisation of the equation of state of dark energy $\omega$ to describe DE dynamical evolution and its impact on the cosmological background as well as its induced effects on the growth of structure \cite{2001IJMPD..10..213C,2003PhRvL..90i1301L}. While at the pure perturbation level, two functions $\mu$ and $\eta$ that alter the relations between the Newtonian and Weyl potential and the matter density contrast were also proposed \cite{Bardeen:1980kt,Mukhanov:1990me,Zhang:2007nk,Malik:2008im,Amendola:2007rr,Pogosian:2010tj}, and alternatively, Effective Field Theory approach (EFT) was also introduced \cite{Bloomfield:2013efa} to define a general action for a broad class of modified gravity theories with effects on the background and the perturbation sector (see \cite{Frusciante:2019xia} and references therein). Finally, more phenomenological inspired parametrisations were considered, such as one for the growth of structure found to be accurately determined by a functional form of the matter density and a growth index $\gamma$ \citep{2005PhRvD..72d3529L}, or another one, $\Sigma$ introduced by \cite{Amendola:2007rr}, which could parameterise the effects of a modified gravity on photon lensing. Here we are interested in constraining three of the aforementioned parameters, i.e. $\omega$, $\gamma$ and $\Sigma$ with a combination of the latest current data. Several have tried to put constrain on one, two, or a combination of the three \cite{Mueller:2016kpu,Hu:2013aqa} without however varying them all together maybe to limit the theoretical biases induced when their phenomenological origin is not treated in a consistent framework.
Here we try to remedy for that by coherently implementing the effect from the three parameters we varied together, $\omega$, $\gamma$ and $\Sigma$ in the cosmological solver \texttt{MGCLASS II} \cite{Sakr:2021ylx} which is a modification to the Boltzmann solver CLASS \cite{2011JCAP...07..034B} to account for the impact of different extensions and modified GR models on the cosmological observables in the quasi-static limit \cite{Mirpoorian:2023utj} or beyond \cite{Baker:2015bva}. This will serve to put, for the first time and within the latest current available data, constraints on these parameters when they are all varied at once, so that to verify whether their allowed values are still compatible with $\Lambda$CDM, but also to prepare the road for exploiting forthcoming data from the next generation stage IV surveys. Often a cosmic evolution for our or other parameterised extensions to $\Lambda$CDM is considered. Here we are also going to compare two chosen ones for $\Sigma$, a constant value and a dynamical one, with the latter inspired by the same way we parameterise the growth with $\gamma$ but using an index we call $\gamma_\ell$ leaving it as constant but free to vary, with the aim, among others, to investigate what would be the effect induced from considering a similar parameterization for both growth and lensing parameters.\\   

The paper is structured as follows: in \autoref{sect:theo} we review the main equations used in the phenomenological approach to test extensions to $\Lambda$CDM model. In \autoref{sect:datacodes} we present the datasets and pipeline used for parameter estimation of the considered models. In \autoref{sect:results}, we show and discuss the results before drawing our conclusions in \autoref{sect:conclusion}

\section{Theoretical context and model implementation}\label{sect:theo}

The evolution of perturbations in modified gravity (MG) could be described by the following relations between the time and scale potentials and the two modified gravity parameterisation $\mu(a,\vec{k})$ and $\Sigma(a,\vec{k})$ in Fourrier space:
\begin{align}
 -k^2\,\Psi(a,\vec{k}) &= \frac{4\pi\,G}{c^2}\,a^2\,\bar\rho(a)\,\Delta(a,\vec{k})\times\mu(a,\vec{k}), \label{eq:mu}\\ 
 -k^2\,\left[\Phi(a,\vec{k})+\Psi(a,\vec{k})\right] &= \frac{8\pi\,G}{c^2}\,a^2\,\bar\rho(a)\,\Delta(a,\vec{k})\times\Sigma(a,\vec{k}), \label{eq:sigma}
\end{align}
where $\bar\rho\Delta=\bar\rho\delta+3(aH/k)(\bar\rho+\bar p)v$ with $\Delta$ the comoving density perturbation of $\delta=(\rho-\bar{\rho})/\bar{\rho}$, and $\bar{\rho}$, $\bar{p}$ and $\bar{v}$ are, respectively, the density, pressure and velocity, with the bar denoting mean quantities. 
$\Phi$ and $\Psi$ are the Bardeen potentials entering the perturbed FLRW metric in flat space, which in the Newtonian gauge is
\begin{equation}
ds^2=a^2\left[-(1+2\Psi)d\tau^2+(1-2\Phi)d\vec{x}^2\right]\,.
\end{equation}
The two functions $\mu(a,\vec{k})$, and $\Sigma(a,\vec{k}))$ encode possible deviations from GR, which is recovered when they are constant and equal to unity. In the following we will consider them as only time dependent.
The implementation of the growth index $\gamma$ here follows the method described in \cite{Pogosian:2010tj}, where however a $\Lambda$CDM background is assumed equivalent to a dark energy equation of state parameter $\omega=-1$. Here we further allow a $\omega \neq-1$.
This will modify the dark energy density $\rho_{\Lambda}$ entering $\Omega_\Lambda$ in the Friedmann-Lema\^{\i}tre equations to 
\begin{equation}\label{equ:rhoDE}
\bar\rho_{\Lambda\left(mod\right)}=\bar\rho_{\Lambda}\left(1+z\right)^{3\left(1+\omega\right)}
,\end{equation}
and propagate into Equ.~\ref{equ:growthmu}, governing the growth of structure, through $H(a)$ and $E_m(a)$.
By combining the system of equations above for modified gravity and the definition of the $\gamma$ parameterization for the growth,  
 \begin{equation}\label{eq:fgrowth}
 \Omega_m(a)^\gamma \, = \frac{d \log D_+}{d \log a}  ,
\end{equation}
where $D_{+} \equiv \Delta(a)_m/a$ is the growth rate, defined in terms of the matter density perturbation $\delta_m$, and $\gamma$ considered constant here (see \cite{Calderon:2019jem,Wen:2023bcj} for more general parameterisations), one can solve the second order equation for scale independent growth 
\begin{equation}
\Delta''(a)+\left[2+{H'(a) \over H(a)} \right]\Delta'(a)-{3 \over 2}{E_m(a) \over E(a)} \mu(a) \Delta(a)=0 ,
\label{equ:growthmu}
\end{equation}
where prime denotes derivative with respect to $\ln a$, $E_m(a)= \Omega_m/a^3$, $E(a)=H^2(a)/H^2_0$ 
 and $\Delta(a)$ the comoving density contrast. After combining with Equ.~\ref{eq:fgrowth} and \ref{equ:rhoDE} to relate $\gamma$, $\omega$ and $\mu(a)$ as:
\begin{equation}
\mu(a)=\frac{2}{3}\Omega_m^{\gamma-1}(a)\left[\Omega_m^{\gamma}(a)+2 -3\gamma + 3(\gamma - \frac{1}{2})\left(\Omega_m(a)+(1+\omega)\Omega_{DE})\right)\right],
\end{equation}
we then end up with two independent parameters $\gamma$ and $\omega$ along with $\Sigma (z)$ which we choose next to be, either constant, or following a parametrization function of the matter density at a given redshift to the power $\gamma_\ell$, 
\begin{equation}
\label{equ:gammaell}
\Sigma (z)= {\Omega_m(z)}^{\gamma_{\ell}},
\end{equation}
inspired from the usually adopted parameterisation function of the growth index,  which we relabel later in the dynamical $\Sigma$ case as $\gamma_f$ to distinguish it from our lensing index.

\section{Datasets used and pipeline}\label{sect:datacodes}

We use CMB temperature, polarization, their cross correlations C$_\ell$ and lensing spectrum D$_\ell$ likelihood \cite{Planck:2019nip} and data released by the Planck satellite mission \cite{Planck:2018vyg} (hereafter Plk18). We also include background observations from BAO measurements \cite{Beutler:2011hx,Ross:2014qpa,BOSS:2016wmc} and combine them with redshift space distortion data (RSD) based on the data set compiled by \cite{Sagredo:2018ahx}. We note that the latter are distinct and not correlated with the BAO data we have chosen. We end by additionally combining with the 3$\times$2pt joint analysis of the galaxy lensing, clustering, and their cross-correlated spectra from the dark energy survey (DES) collaboration \cite{DES:2017myr,DES:2021wwk} where we limit our analysis to the linear scales, i.e. we follow \cite{Ade:2015rim} and \cite{Abbott:2018xao,DES:2022ccp} method and consider the difference between the nonlinear and linear-theory predictions in the standard model at best-fit values of the cosmological parameters and retain only the data points that contributes the least to increase this quantity by adopting a conservative cut (see appendix B of \cite{Zucca:2019xhg} for more details on that) going below the `standard' threshold such that scales in the linear regime are kept practically.
We run our MCMC using \texttt{MGCLASS II} \cite{Sakr:2021ylx} \footnote{\url{https://gitlab.com/zizgitlab/mgclass--ii}} which is interfaced with the cosmological data analysis code \texttt{MontePython} \cite{2013JCAP...02..001A} in which the RSD code was first included by \cite{Arjona:2020yum} while the DES likelihood was implemented in it by us based on the official public one. We varied the following cosmological parameters with the flat priors $(\Omega_b,\Omega_M,h\footnote{$h= H_0 /100 (\rm km.\rm s^{-1}.\rm Mpc^{-1}$}, n_s, \sigma_8 )=(0.03 - 0.05,\, 0.1 - 0.9,\, 0.3-1.2,\, 0.7-1.0, \, 0.5 -1.2)$ , along with our extension to the standard model parameters ($\gamma_f, \gamma_\ell \, {\rm or} \, \Sigma, \omega$) with non informative priors, while we kept the same settings and priors for the DES likelihood as detailed in Table I of \cite{DES:2017myr}.

\section{Results}\label{sect:results}

\begin{figure}
\centering
\includegraphics[width=1.0\textwidth]{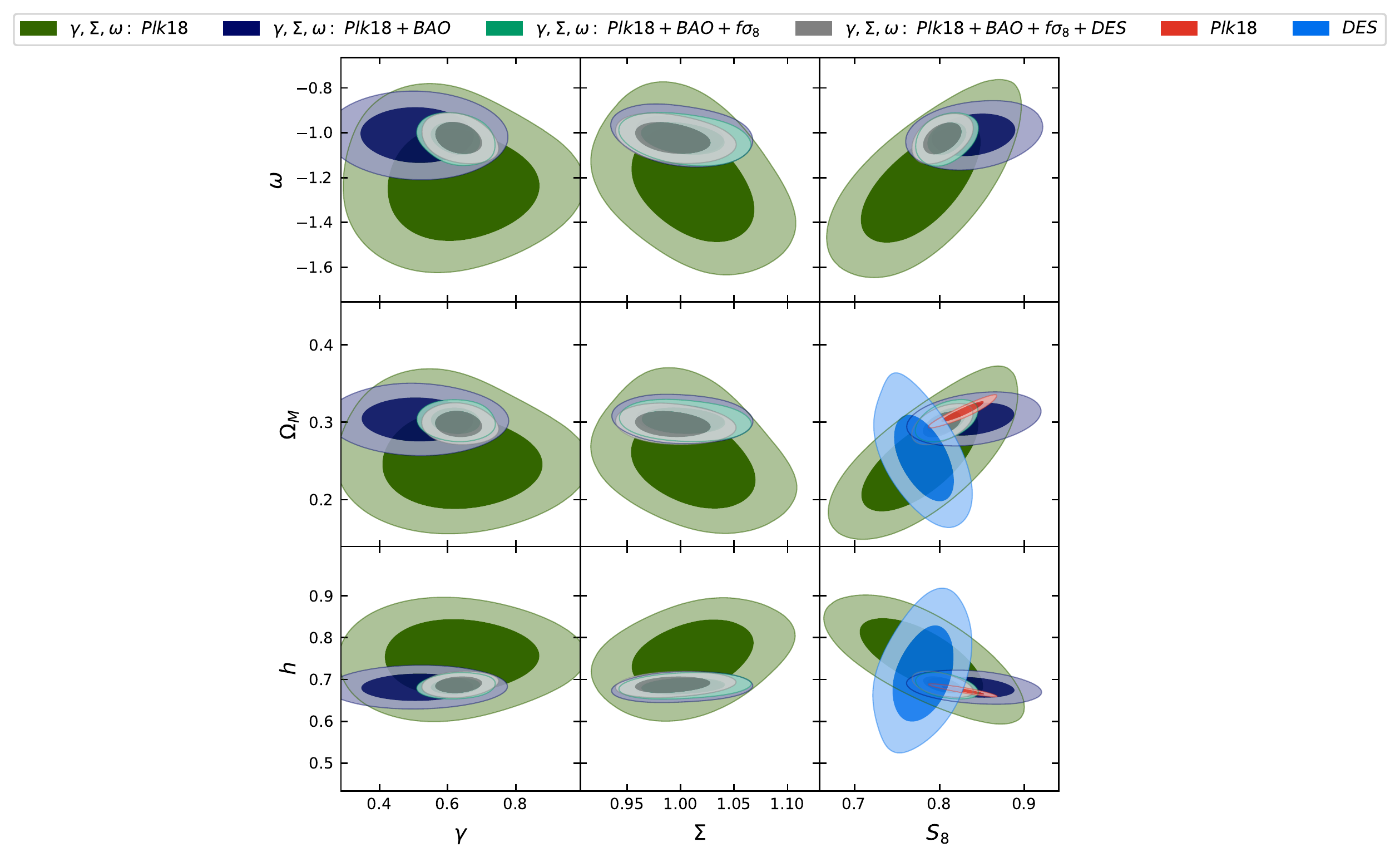}
\caption{68\% and 95\% confidence contours for $\Omega_m$, $h$, $\omega$, $S_8$, $\gamma$ and $\Sigma$ inferred from different combinations of: CMB $C_{\ell}^{TT,TE,EE}$ and its lensing potential from Plk18, BAO observations, $f\sigma_8$ measurements and 3x2pt photometric lensing and galaxy correlations and cross correlations from DES survey.}
\label{fig1}
\end{figure}

\begin{figure}
\centering
\includegraphics[width=1.0\textwidth]{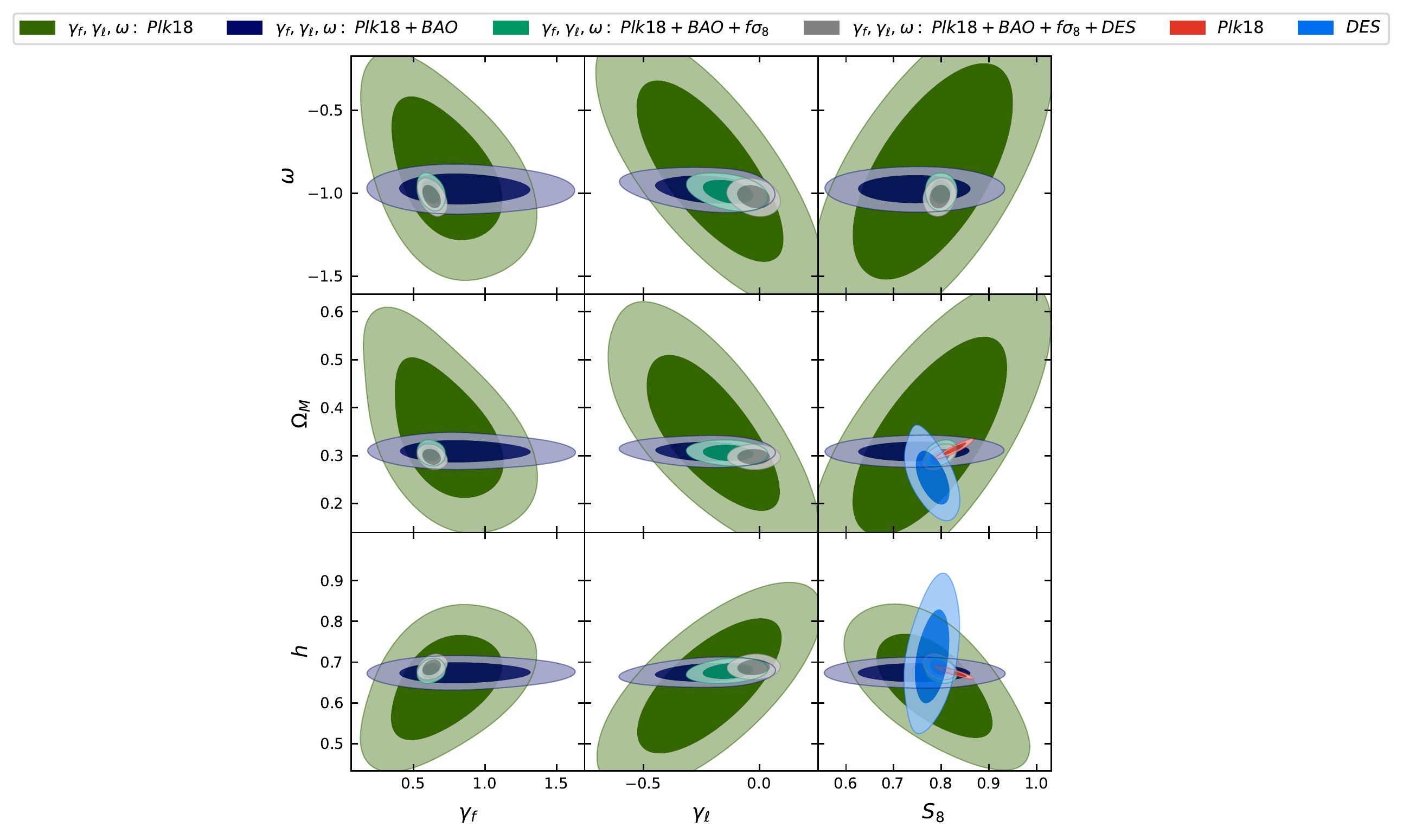}
\caption{68\% and 95\% confidence contours for $\Omega_m$, $h$, $\omega$, $S_8$, $\gamma_f$ and $\gamma_\ell$ from different combinations of: CMB $C_{\ell}^{TT,TE,EE}$ and its lensing potential from Plk18, BAO observations, $f\sigma_8$ measurements and 3x2pt photometric lensing and galaxy correlations and cross correlations from DES survey.}
\label{fig2}
\end{figure}

In Fig.~\ref{fig1} and \ref{fig2} we show constraints on $\Omega_m$, $h$, $\omega$, $\gamma$, $\sigma$ and $S_8={(\Omega_m/0.3)}^{0.5}\sigma_8$, obtained after varying all the cosmological and nuisance parameters of the probes mentioned in \autoref{sect:theo} and that for two cases: one with a constant $\Sigma$ and the other with our new parameterization $\Sigma(z) = {\Omega_m(z)}^{\gamma_\ell}$. For each case, we start by showing constraints from CMB $C_{\ell}^{TT,TE,EE}$ and lensing potential from Plk18, and continue by adding BAO first, then $f\sigma_8$ growth measurements next, to finish with DES 3$\times$2pt joint analysis of the galaxy clustering and lensing from photometric redshift probe. This order is motivated by what each probe is expected to add as additional constraints on our three beyond $\Lambda$CDM parameters. As so, the CMB temperature and polarisation angular power spectrum and its extracted lensing quadrupole probe will start by putting constraints on some or all of our beyond $\Lambda$CDM parameters as seen in \cite{Ade:2015rim} and \cite{Planck:2018vyg} as well as \cite{Sakr:2021ylx}. Next BAO is expected to additionally constrain notably $\omega$ without much effect on the growth and lensing parameters, since the latter will be respectively constrained by the addition of the redshift space distortions data and the lensing from DES data. The latter with the galaxy clustering data will especially constrain the growth and lensing MG parameters from the leverage gained from observations covering $0<z<1.5$, combined with those coming from higher redshifts from CMB lensing power spectrum that are mainly sensitive to large scale structure around $z \sim 2$. For the two models, the constant or dynamical $\Sigma$ one, we also show separately, $\Lambda$CDM constraints from Plk18 data and DES 3$\times$2pt data to get better insights about how the constraints are changing with respect to the standard cosmological model.
For the two models, either $\Sigma$ constant or parameterised through $\gamma_\ell$, we observe that the Plk18 data allow much freedom for all of the triggering parameters (green lines) though still compatible with their equivalent $\Lambda$CDM values. We also notice that the constraints on $S_8$ and $h$ are loose so that the range covers the values subject of tensions between local and deep probes on these two parameter bounds (see \cite{Perivolaropoulos:2021jda,DiValentino:2021izs,Sakr:2021jya} for more on this subject). In both parameterisations of $\Sigma$, we especially see $\omega$ preferring high negative values as was the case in similar studies \cite{DES:2017myr,DES:2021wwk} though it is showing stronger correlations with the other two beyond $\Lambda$CDM parameters for the model with dynamical $\Sigma$.
As expected, adding BAO data strongly constrains $\omega$ for both models (dark blue lines) as well as $h$ and $\Omega_m$ while having a small effect on the growth index $\gamma$. For the MG lensing parameter, we see also that the gain is small for $\Sigma$ constant but is more substantial for $\gamma_\ell$. This is due to the fact that we have broken the degeneracy on the latter when we strongly constrained other cosmological parameters especially $\Omega_m$ since now we are limiting  the freedom of $\Sigma = \Omega_m(z)^{\gamma_\ell}$ to vary. We also observe that all the parameters are still compatible with their equivalent $\Lambda$CDM values though the growth index $\gamma$ is showing preference for values below 0.55 in the constant $\Sigma$ case and the lensing index $\gamma_\ell$ is preferring negative values for the dynamical $\Sigma$ case. The reason for BAO not equally breaking degeneracy between matter density and the growth index as was the case for the lensing index is that $\gamma$ has very little effect on the CMB perturbations at the recombination epoch where $\Omega_m(z)$ value is at the level of unity.
The situation substantially changes when we add growth data (light green lines), which mainly put strong constraints on the growth index $\gamma$ for both models, while $\Sigma$ constraints are almost unchanged in the first model and tighter for the second even if they are still preferring negative values. This comes from the fact that $f\sigma_8$ measurements, by constraining $\gamma$ are also doing the same for $\Omega_m$ breaking its degeneracy with $\gamma_\ell$. We notice also that $S_8$ is back in being in slight tension with its $\Lambda$CDM value, a fact later consolidated when we additionally add DES datasets. In the latter case, we see (gray lines) that DES data has not added much improvement to the growth index $\gamma$ bounds in both cases although we see that the latter is now well constrained around 0.6 not far from its fiducial $\Lambda$CDM value. While for the MG lensing parameter $\Sigma$, we observe that the gain in the constant case is small with respect to what we already had from Plk18+BAO+RSD. However, the improvement is noticeable for the model with dynamical $\Sigma$,  previously showing preferences for negative values but now shifted to be well constrained around the null value corresponding to $\Lambda$CDM and GR. We show in table~\ref{tab:wparameter_values} and table~\ref{tab:ellparameter_values} at the level of 68\% confidence the bounds obtained by the different combinations in the two cases on the beyond $\Lambda$CDM parameters considered in this work. The values in all cases but especially the CMB+BAO one agree with the recent work of \cite{Specogna:2023nkq} where they limit to these probes while the most constraining case CMB+BAO+$f\sigma8$+DES agrees well with another study using all these probes \cite{Nguyen:2023fip} albeit the fact that they followed instead the method of rescaling the power spectrum for a new value of $\gamma$, and that they did only consider one extra parameter $\gamma$. The latter agreement is then due to the fact that in the most constraining case $\omega$ and $\Sigma$ are close to their $\Lambda$CDM values. \\
\begin{table}
\centering
\begin{tabular}{c|c c c c|}
\cline{2-5}  
  &
  CMB &
  CMB + BAO &
  CMB + BAO + $f\sigma8$ &
  CMB + BAO + $f\sigma_8$ + DES \cr
   \hline
\multicolumn{1}	{|c|}	{$\gamma	$}	&	0.63$\pm$0.11 &	0.518$\pm$0.072	&	0.624$\pm$0.046	&	0.633$\pm$0.044		\\	\hline
\multicolumn{1}	{|c|}	{$\omega	$}	&	-1.23$\pm$0.15 &	-1.009$\pm$0.053 &	-1.025$\pm$0.047	&	-1.025$\pm$0.045		\\	\hline
\multicolumn{1}	{|c|}	{$\Sigma	$}	&	1.011$\pm$0.032 &	0.998$\pm$0.025	&	1.002$\pm$0.024	&	0.992$\pm$0.022		\\	\hline
\end{tabular}
\caption{constraints at the 68\%  level of $\omega$, $\gamma$ and $\Sigma$ inferred from different combinations of: CMB $C_{\ell}^{TT,TE,EE}$ and its lensing potential from Plk18, BAO observations, $f\sigma_8$ measurements and 3x2pt photometric lensing and galaxy correlations and cross correlations from DES survey.}
\label{tab:wparameter_values}
\end{table}

\begin{figure}
\centering
\includegraphics[width=0.6\textwidth]{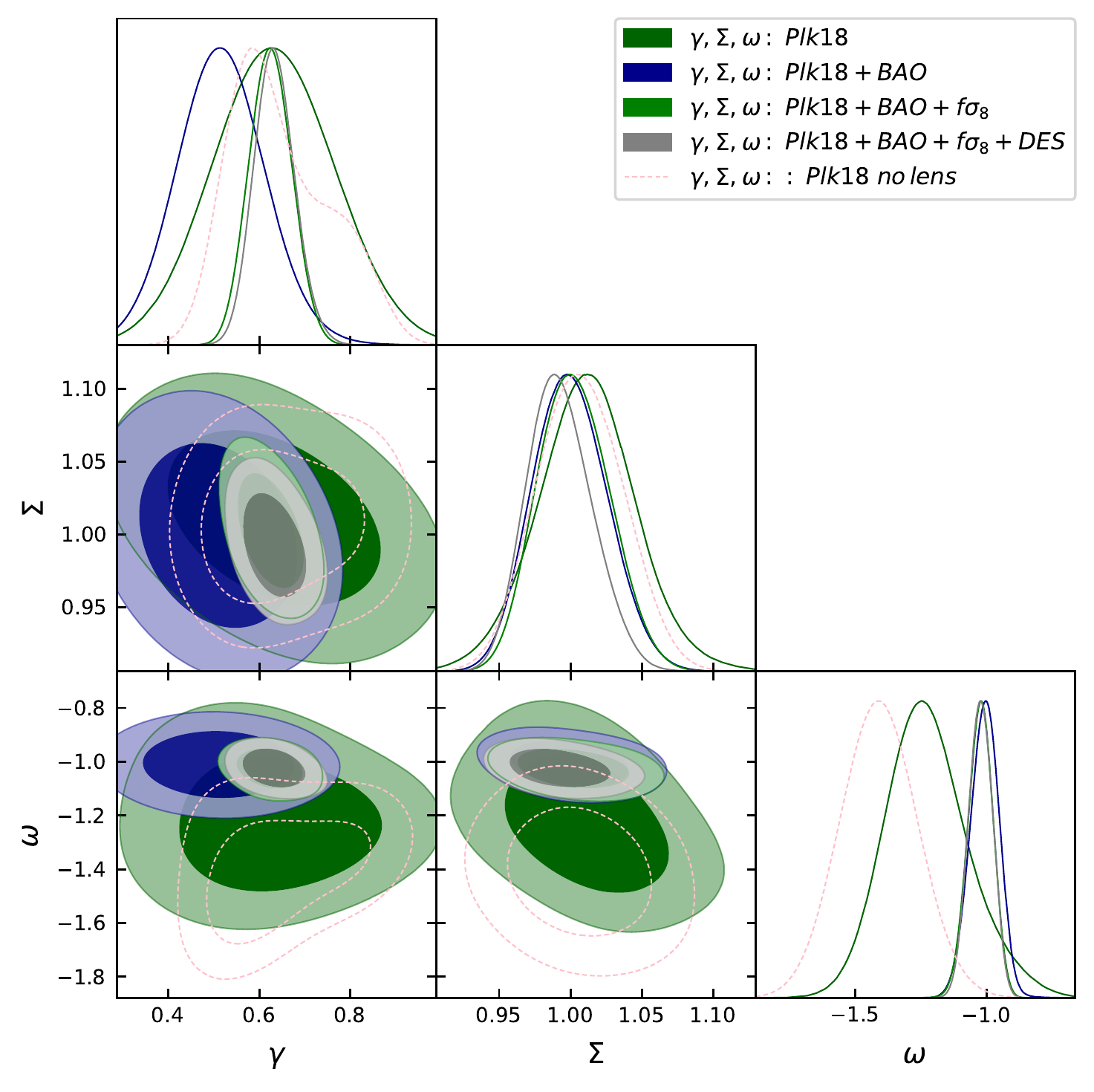}
\caption{68\% and 95\% confidence contours for $\gamma$, $\Sigma$ and $\omega$, inferred from different combinations of: CMB $C_{\ell}^{TT,TE,EE}$ and its lensing potential from Plk18, BAO observations, $f\sigma_8$ measurements and 3x2pt photometric lensing and galaxy correlations and cross correlations from DES survey; in comparison to the same combinations but using CMB $C_{\ell}^{TT,TE,EE}$ from Plk18 alone.}
\label{ellvsnoellgamsigwall}
\end{figure}

\begin{figure}
\centering
\includegraphics[width=0.6\textwidth]{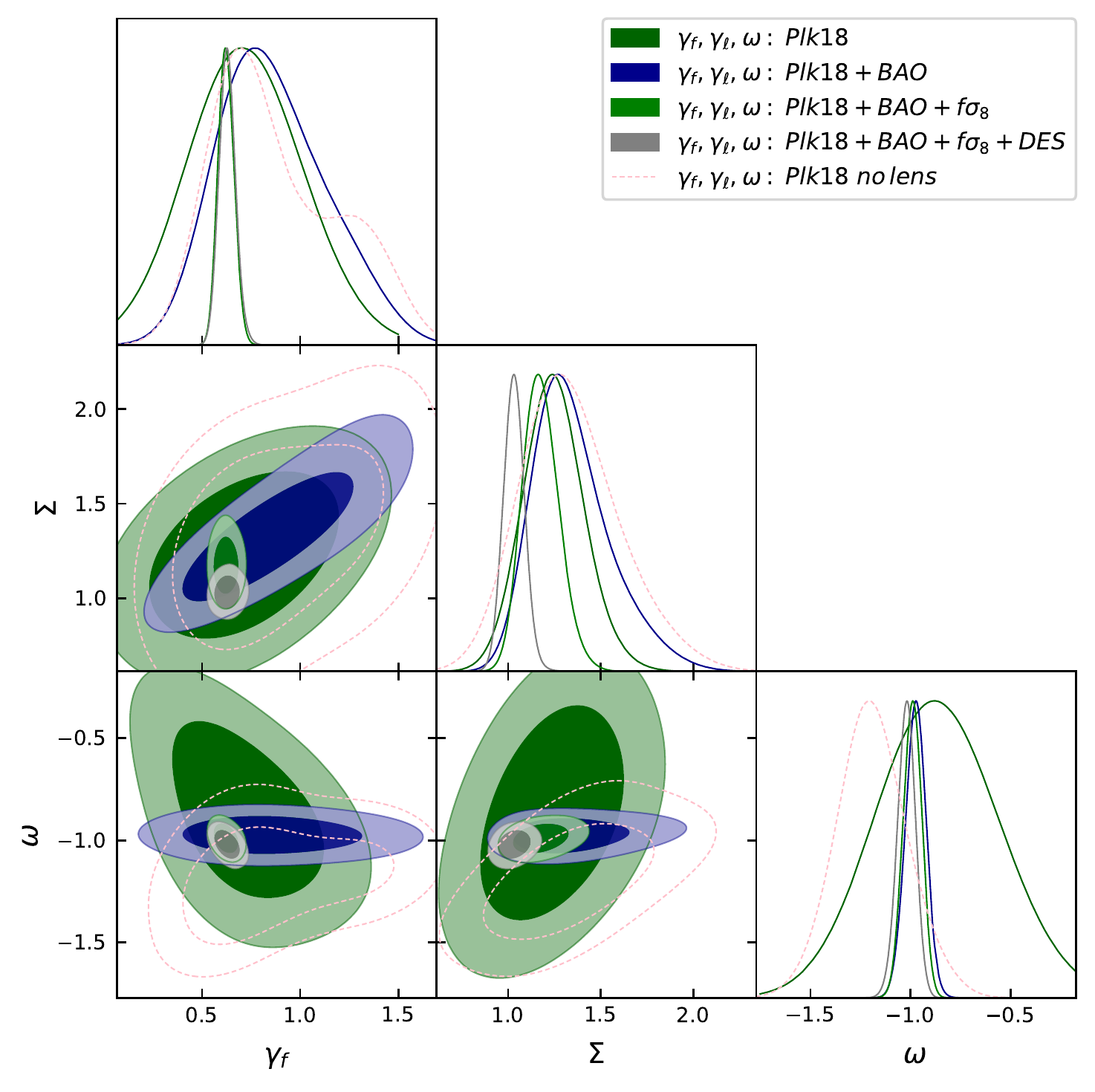}
\caption{68\% and 95\% confidence contours for $\gamma_f$, $\omega$ and $\Sigma$ derived from $\gamma_\ell$ following Equ.~\ref{equ:gammaell} inferred from different combinations of: CMB $C_{\ell}^{TT,TE,EE}$ and its lensing potential from Plk18, BAO observations, $f\sigma_8$ measurements and 3x2pt photometric lensing and galaxy correlations and cross correlations from DES survey; in comparison to the same combinations but using CMB $C_{\ell}^{TT,TE,EE}$ from Plk18 alone.}
\label{ellvsnoellgamfgamsigal}
\end{figure}

These results could be further understood when looking at Fig.~\ref{ellvsnoellgamsigwall} and Fig.~\ref{ellvsnoellgamfgamsigal} where we show contours for the two models with additional plots obtained from Plk18 data without including its lensing probe in order to emphasise on the evolution of the degeneracies and correlations between our beyond $\Lambda$CDM parameters.  We also show the derived parameter $\Sigma$ instead of $\gamma_\ell$ to allow a better comparison between the two cases.
We see first in Fig.~\ref{ellvsnoellgamsigwall}, with the constant $\Sigma$ case, that there are no differences in the constraints on $\gamma$ or $\Sigma$ coming from the Plk18 lensing. The difference is seen however on those on $\omega$, further favouring negative values when using Plk18 without the lensing data, a result due to the fact that CMB lensing potential is affected by the large scale structures at $z\sim2$ where $\omega$ effects on their formation are noticeable. We also see a correlation and degeneracy between $\Sigma$ and $\gamma$ that is not broken by the BAO data alone but rather when we add growth measurements and DES data. We also observe a correlation between $\Sigma$ and $\omega$ which is not present between $\omega$ and $\gamma$ at the Plk18 only level  $\Omega_m$ value is close to one as mentioned earlier. This could explain why BAO was able of constraining also $\Sigma$ so well  because it does the same to the related correlated parameter $\omega$.
While for the case where $\Sigma$ is parameterised function of $\gamma_\ell$, we observe in Fig.~\ref{ellvsnoellgamfgamsigal} the same behaviour in the  $\Sigma - \omega$  contours but a different one for the $\Sigma - \gamma$ contours. The latter are showing positive correlations especially when using Plk18 data alone since now for both parameters the constrains are driven by $\Omega_m$ that is less constrained by CMB at the recombination epoch. Both parameters are also not correlated with $\omega$ which explains why BAO have not the same effect we saw in the $\Sigma$ constant case. We also understand why $f\sigma_8$ measurements were able of strongly constraining $\Sigma$ even before further consolidation from DES, since their addition breaks the degeneracy with $\gamma$ allowing us to benefit from BAO and CMB constraints on the matter density parameter  to tighten those on both $\Sigma$ and $\gamma$ parameters.
\begin{table}
\centering
\begin{tabular}{c|c c c c|}
\cline{2-5}  
  &
  CMB &
  CMB, BAO &
  CMB, BAO, $f\sigma8$ &
  CMB, BAO, $f\sigma_8$ + DES \cr
   \hline
\multicolumn{1}	{|c|}	{$\gamma_f	$}	&	0.72$\pm$0.22	&	0.86$\pm$0.27&	0.622$\pm$0.040	&	0.628$\pm$0.040		\\	\hline
\multicolumn{1}	{|c|}	{$\omega	$}	&	-0.87$\pm$0.24	&	-0.975$\pm$0.051	&	-0.991$\pm$0.046	&	-1.021$\pm$0.045		\\	\hline
\multicolumn{1}	{|c|}	{$\gamma_\ell	$}	&	-0.22$\pm$0.13	&	-0.24$\pm$0.13 &	-0.135$\pm$0.070	&	-0.025$\pm$0.045		\\	\hline
\end{tabular}
\caption{constraints at the 68\% level of $\gamma_f$, $\omega$ and $\gamma_\ell$ inferred from different combinations of: CMB $C_{\ell}^{TT,TE,EE}$ and its lensing potential from Plk18, BAO observations, $f\sigma_8$ measurements and 3x2pt photometric lensing and galaxy correlations and cross correlations from DES survey.}
\label{tab:ellparameter_values}
\end{table}

\section{Conclusion}\label{sect:conclusion}
In this work we performed a Bayesian analysis to obtain simultaneous constraints on three parameters commonly used each alone to test deviations from $\Lambda$CDM, namely the dark energy equation of state parameter $\omega$, the growth index $\gamma$ and the Weyl potential parameter $\Sigma$ and that in a consistent way within the same Boltzmann and cosmological solver used to compute the different theoretical observables within the new cosmologies.
The analysis was conducted using different combinations of CMB $C_{\ell}^{TT,TE,EE}$ and lensing potential, BAO observations, $f\sigma_8$ measurements and 3$\times$2pt joint analysis of the galaxy and lensing clustering in photometric redshifts from DES survey, and that by considering two parameterisation for $\Sigma$, first as a constant and second following a similar approach usually performed for the growth index $\gamma_f$, by formulating  $\Sigma$ as function of the matter density at a certain redshift elevated to a lensing index $\gamma_\ell$ introduced for the first time in this work.
For the case of a constant $\Sigma$, we found that the CMB data allows deviations from $\Lambda$CDM though still compatible with the latter, while adding BAO strongly constrains the $\omega$ parameter from its impact on the background evolution, permitting us to a lesser extent to constrain the lensing parameter $\Sigma$  as a result of breaking the degeneracies between the latter and the other cosmological parameters such as $\Omega_m$ and $h$. Adding the growth $f\sigma_8$ measurements and the clustering and lensing correlations of galaxies from photometric redshifts then strongly limits the bounds on $\Sigma$ and those on the growth index despite a shift to the latter values to within 2$\sigma$ from its $\Lambda$CDM ones. For the case where we parameterise $\Sigma$ as function of the matter density and a lensing index $\gamma_\ell$, we found strong degeneracies between the aforementioned parameters due to the common presence of matter density ingredient when constrained only by CMB data. The addition of BAO was only able of breaking degeneracies with $\omega$ and a further combination with the $f\sigma_8$ data was needed to constrain the growth index $\gamma_f$ to within 2$\sigma$ from its fiducial value. Still the lensing index $\gamma_\ell$ was showing preference for negative values even though compatible with $\Lambda$CDM, but that was the case until the clustering and lensing correlations of galaxies from photometric redshifts were further added, bringing it back to having a maximum likelihood close to its equivalent $\Lambda$CDM value. 
We conclude that $\Lambda$CDM is still compatible with a combination of the large cosmological datasets used in this work despite the three phenomenological model extensions adopted and a mild tension for $\gamma$, while adopting a similar parameterisation for the lensing and growth of structure function of the matter density, offers room for substantial deviation from general relativity, especially for the Weyl potential parameter, which was not strongly constrained  in this case until we combined all of the datasets used in this work.

\acknowledgments

Z.S. acknowledges funding from DFG project 456622116 and support from the IRAP Toulouse and IN2P3 Lyon computing centers.

\bibliographystyle{unsrt}
\bibliography{references}

\end{document}